\documentstyle[twoside,fleqn,espcrc2]{article}

\newcommand{\AmS}{{\protect\the\textfont2
  A\kern-.1667em\lower.5ex\hbox{M}\kern-.125emS}}

\begin{document}

\title{Direct $CP$ Violation in a Variety of  $B$ Decays}

\author{George W. S. Hou
 \address{Department of Physics, National Taiwan University, \\
                   Taipei, Taiwan 10764, R.O.C.}%
   \thanks{Supported by grant NSC 83-0208-M-002-023 of the Republic of China.
           Work done in part in collaboration with J.-M. G\' erard
                                          and with I. Dunietz.}
}

\begin{abstract}
We give a brief review of direct $CP$ violation effects
in charmless $B$ decays,
both at the inclusive and at the exclusive level.
It is emphasized that typical exclusive final states
are composed of 2--4 $K$ or $\pi$'s,
hence particle identification
(in terms of $K/\pi$ separation) is of crucial
importance for studying $CP$ violation in these modes.
Another kind of  ``charmless" final state is also emphasized:
$b\to dc\bar c$. The decay rate is tree dominant, while ``$q^2$"
is necessarily above $c\bar c$ threshold
hence in the domain of perturbative QCD.
The $CP$ asymmetry is therefore relatively clean and reliable.
Theoretically, a few $\times 10^7$ $B$'s would suffice
if one could accumulate modes like
$B^+\to D^{+\left(*\right)} \bar D^{0\left(*\right)}$.
\end{abstract}

\maketitle
\pagestyle{plain}

\section{Mechanism}

The essence is two interfering amplitudes
with two different kinds of phases.

Decay rate asymmetry between particle {\it vs.} antiparticle,
\begin{equation}
  a_{CP} \equiv {\Gamma(B \to f) - \overline{\Gamma}(\overline{B} \to \bar f)
\over
                 \Gamma(B \to f) + \overline{\Gamma}(\overline{B} \to \bar f)},
\end{equation}
provides a measure of $CP$ violation.
However,
$CPT$ invariance {\it demands} the equality of the {\it total} rates,
\begin{equation}
 \Gamma_{\rm tot} = \overline{\Gamma}_{\rm tot}.
\end{equation}
Unitarity of the $S$-matrix then demands there be
final state rescattering,
or, presence of {\it absorptive parts}, as a necessary condition
for $CP$ violating rate asymmetries.

Define
\begin{equation}
{\cal M} = {\cal D} - \frac{i}{2}{\cal A},
\end{equation}
where ${\cal D}$ and ${\cal A}$ are the dispersive
and absorptive (on-shell rescattering) amplitudes, respectively.
With physical $CP$ violating phases, such as KM (Kobayashi--Maskawa)
model with 3 or more quark generations,
${\cal D}$ and ${\cal A}$ can both be complex.
Hence, $\overline{\cal M} = {\cal D}^* - \frac{i}{2}{\cal A}^*$
for the process involving antiparticles,
and we find
\begin{equation}
\vert{\cal M}\vert^2 = \vert{\cal D}\vert^2
                                    + \frac{1}{4}\, \vert{\cal A}\vert^2
                                    + {\rm Im}\, ({\cal D}^*{\cal A}),
\end{equation}
while for $\vert\overline{\cal M}\vert^2$,
the ${\cal D}$-${\cal A}$ interference term changes sign.
Suppressing phase space integration, {\it etc.} for simplicity,
we have
\begin{equation}
a_{CP}  = {{\rm Im}\, ({\cal D}^*{\cal A})  \over
                  \vert{\cal D}\vert^2  + \frac{1}{4}\, \vert{\cal A}\vert^2}.
\end{equation}
It is therefore clear that,
if  ${\cal A} = 0$ (no absorptive part),
or if  ${\cal D}$ and ${\cal A}$ have {\it same} phase,
then there can be no observable $CP$ violating asymmetry.

\subsection{Mixing Induced}

$B^0$-$\overline{B}^0$ mixing provides a fascinating vehicle
for the above mechanism.
Here, the dispersive part comes from direct  decay,
$\langle f\vert {\cal H}_W \vert B^0\rangle$, where $f$ is usually
chosen as an eigenstate of $CP$, while
the absorptive part comes about  via
$\langle f\vert {\cal H}_W \vert \overline{B}^0\rangle
\langle \overline{B}^0\vert \mbox{4-Fermi}\, \vert B^0\rangle$,
{\it i.e.} mixing ($B^0\to \overline{B}^0$)
followed by decay of $\overline{B}^0$
to $f$.
The two amplitudes clearly could have different $CP$ violating phases.
In other words, if one starts with $B^0$ at $t = 0$,
the state could decay via ${\cal M}(B^0(t)\to f)$
or via $\overline{\cal M}(\overline{B}^0(t)\to f$) at time $t$,
assuming $f$ is $CP$ eigenstate.
{\it The time dependent oscillation phase $e^{i\Delta m\, t}$
serves the function as absorptive part!}
This is why mixing dependent $CP$ asymmetries are so cute
(quantum mechanics!)
and clean: $\Delta m$ is already measured and known,
while the oscillation phase $e^{i\Delta m\, t}$
is to be measured directly,
presumably by silicon based micro vertex detectors.

The best known and best studied mode is
$B^0\to \psi K_S$, which would measure, without any hadronic ``pollution",
the angle $\beta$ ($\phi_1$) of the 3 generaton unitarity triangle.
The mode has been under intense simulation study, for asymmetric
$B$ Factories at $\Upsilon (4S)$ energies,
and for hadronic machines as well,
both for fixed target and collider modes.
Note that this mode has already been seen by CDF since 1990.
With SVX installed, if  the so-called
``$B^{**}$" or  ``bachelor pion"
tagging-via-production method works,
there are some hopes that one may
observe $CP$ violation in this mode at the Tevatron collider,
with Main Injector upgrade.
Thus, this mode has been the conventional bet, but,
as we have seen, one {\it needs 1) TAGGING, 2) two vertices}
(for sake of $t_1 - t_2$ measurement).
And hereby lies the challenge for these type of modes.
It should also be remembered that once one  goes beyond this
mode, the challenge increases.

\subsection{Direct (in decay)}

The two interfering amplitudes are now
$\langle f\vert {\cal H} \vert B^0\rangle$
and
$\langle f\vert {\cal H}_{\rm strong} \vert i\rangle
\langle i\vert {\cal H}_W\, \vert B^0\rangle$,
where $i$ is any possible decay final state that could rescatter to
the particular final state $f$ (not necessarily $CP$ eigenstate).
The effective Hamiltonian of the first term
includes the so-called penguin operators, induced by
an interplay of weak and strong interactions.
For the second term, the physical processes are, for example
$B\rightarrow K + n\, \pi,\ D\bar DK+ n\,\pi$ {\it etc.},
rescattering via strong interactions into final states such as
$K + n\,\pi$, $K\phi$ {\it etc.}.
The mechanism was first applied to $B$ system
by Bander, Silverman and Soni (BSS) \cite{BSS79}.
The cute point is that strong interaction rescattering
in the perturbative regime is readily incorporated in
penguin absorptive parts via perturbative QCD, and by
assuming duality, namely, that quark-gluon picture
corresponds to the sum over all possible hadronic states.
However, for the experimentally more tangible exclusive processes,
both perturbative and {\it nonperturbative}
strong phases enter, making them rather ``unclean".
This is why these type of modes, in contrast to mixing-dependent ones,
have received scant conventional bets.

These modes should not be overlooked, however, since they
could be {\it discovery} modes.
The main advantage is that they are typically self-tagging
and could be observed for both $B^+$ and $B^0$ decays.
A symmetric $B$ Factory would suffice.
Besides theoretical uncertainties, the main experimental challenge
is to log the data down (trigger problem for hadronic machines),
and have good (charge and) $K/\pi$ identification.

\section{Direct $CP$ Violation}

Since there has been much devoted work and discussions
in the past already, the purpose here is to summarize the salient features
and point out theoretical and experimental problems/challenges.

\subsection{Inclusive $b\to s$ and $b\to d$}

It was by studying inclusive charmless $B$ decays
that G\' erard and Hou \cite{GH89} uncovered the full importance
of $CPT$ invariance and unitarity.

At the tree level, one has only
$b\to su\bar u$ (one could always replace the $s$ by $d$).
There can be no $CP$ asymmetries because there is only one amplitude
and there is no rescattering.
At ${\cal O}\, (\alpha_s)$,
the tree level (dispersive) $b\to su\bar u$ amplitude
can interfere with
\begin{equation}
\overbrace{b\longrightarrow sc\bar c}^{\mbox{ Weak}}
\underbrace{\to sg^*
                     \to}_{\mbox{Strong}}
su\bar u
\end{equation}
which is absorptive. The physical  $sc\bar c$ final state
could rescatter inelastically via strong interactions.
The beauty is, assuming the validity of perturbative QCD,
this rescattering amplitude is given by calculating
the absorptive part of the one-loop penguin amplitude.
The validity of perturbative QCD is justified
by the fact that the $c\bar c$ has to disappear
into $u\bar u$, hence the $q^2$ of the
$u\bar u$ pair is necessarily above $4m_c^2$.

\begin{table*}[hbt]
\setlength{\tabcolsep}{1.5pc}
\newlength{\digitwidth} \settowidth{\digitwidth}{\rm 0}
\catcode`?=\active \def?{\kern\digitwidth}
\caption{
(Semi-)inclusive BR and asymmetries for $b \to s$
and $b \to d$ processes,
for $\rho = -0.5$, $\eta = 0.15$ and $m_t = M_W$.
The dependence on $m_t$ is weak, while dependence on
$\rho$ is more involved,
but the asymmetries roughly  scale linearly with $\eta$.
Hence, if $\eta = 0.3$, the asymmetries are typically twice as large.
${\it a}$(${\it a_o}$) is with(out) the extra
$\alpha_s^0\cdot\alpha_s^2$ interference term taken into account.
``$0.0$" stands for a very small positive number.
}
\begin{tabular*}{\textwidth}{@{}l@{\extracolsep{\fill}}ccc}
\hline
                &  \multicolumn{1}{c}{BR (\%)}
                &  \multicolumn{1}{c}{$a_0$ (\%)}
                &  \multicolumn{1}{c}{$a$ (\%)}    \\
\hline
$b\to su\bar u$ &    0.46   &    1.2    &   0.0   \\
$b\to sd\bar d +  ss\bar s$ &    0.54   &   0.5   &   0.5  \\
Total Charmless $b\to s$    &    1.19   &   0.7   &   0.2  \\
\hline
$b\to du\bar u$ &    0.71   &   -0.7    &  -0.0   \\
$b\to dd\bar d +  ds\bar s$ &    0.07   &  -4.2   &  -4.2  \\
Total Charmless $b\to d$   &    0.80   &  -1.0   &  -0.4  \\
\hline
$b\to dc\bar c$ &    0.75   &    0.85    &   same   \\
\hline
\multicolumn{4}{@{}p{120mm}}{Adapted from: J.-M. G\' erard and W.S. Hou,
                                           Phys. Rev. D {\bf 43}, 2909 (1991)
\cite{GH91a}.}
\end{tabular*}
\end{table*}

The virtual gluon, however, can go into $d\bar d$ and
$s\bar s$ as well. Furthermore, the penguin amplitude
itself has dispersive contributions that
mediate $b\to sq\bar q$ transitions.
It turns out that,
because the dominant $c\bar c$ and $t\bar t$ (virtual) states
are not Cabibbo suppressed compared to usual $b$ decay, while
tree level $b\to su\bar u$ is multiply Cabibbo suppressed,
the loop-induced penguin amplitude is stronger than
the tree level dispersive amplitude in strength \cite{GPR81}.
One therefore has an incentive to check and see whether
the penguin decay by itself could exhibit $CP$ asymmetries.
Since the penguin process possesses both dispersive and absorptive
contributions, and since all three generations contribute via
this loop process, indeed, penguin--penguin interference
could lead to $CP$ violating asymmetries in pure penguin rates \cite{BSS79}.
This, however, occurs at ${\cal O}\, (\alpha_s^2)$.
Hereby lies the subtlety.
Because of demands of $CPT$/unitarity, and, perhaps more important,
the proximity of $m_b$ to $c\bar c$ threshold,
it turns out \cite{GH89} that one not only has to include the
effectively two-loop, ${\cal O}\, (\alpha_s^2)$
penguin--penguin interference $CP$ violation effect,
one has to incoporate another effectively two-loop,
${\cal O}\, (\alpha_s^2)$ effect due to
tree--penguin interference where one has a
cut on the virtual gluon line
(absorptive part of gluon self-energy).
In effect, the ${\cal O}\, (\alpha_s^2)$ term
overwhelms the ${\cal O}\, (\alpha_s)$ term,
which is counter-intuitive but is a truly subtle effect.
For a detailed discussion, see ref. \cite{GH91a}.

It suffices for our present purpose to see the numerical
impact of the above discussion,
which we display in Table 1.
The $b\to dc\bar c$ mode will be discussed in Sec. 3.
Experimentally it remains to be seen whether one can
sum over modes to study inclusive asymmetries in
a realistic way.

Note that since the $a_{CP}$ in pure penguin modes
$b\to sd\bar d,\ ss\bar s$ are already at ${\cal O}\, (\alpha_s^2)$,
they are unaffected by the additional ${\cal O}\, (\alpha_s^2)$
tree--penguin interference. However, admittedly, the effect
of incorporating this extra ${\cal O}\, (\alpha_s^2)$ term
on the $b\to su\bar u$ mode is rather dramatic.
The fact that the effect is similar for $b\to d$ transitions
means that the mechanism does not depend on KM structure.
One other point: $b\to d$ type penguins possess much larger
$a_{CP}$'s than $b\to s$ type penguins.
This would carry over to the exclusive case, to which we now turn.

\subsection{Exclusive Modes: Model Dependence}

The problem here lies with theory:
One needs to calculate hadronic matrix elements of
4-quark operators, just to evaluate decay rates.
Despite advances due to Heavy Quark Effective Theory,
because charmless final states always involve light hadrons,
this is still something of an art form.
We shall use the so-called BSW (Bauer--Stech--Wirbel) model.

\begin{table*}[hbt]
\setlength{\tabcolsep}{1.5pc}
 \settowidth{\digitwidth}{\rm 0}
\catcode`?=\active \def?{\kern\digitwidth}
\caption{
Summary table for penguin-related exclusive modes (illustrative) of interest.
}
\begin{tabular*}{\textwidth}{@{}l@{\extracolsep{\fill}}lcc}
\hline
         {Inclusive (quark)}
      &  {Exclusive (hadron)}
      &  {BR}
      &  {$a_{CP}$}    \\
\hline
{\boldmath $b\to su\overline u$}
                          &    $B^0\to K^+\pi^-,\ K^{+*}\pi^-$
                          &
                          &    $\sim -2\%$ \ -- \ $+ 2\%$   \\
\{tree $\oplus$ penguin\}
                          &    $B^+\to K^+\rho^0,\ K^{0*}\pi^+$
                          &    ${\cal O}(10^{-5})$
                          &    (perturbative phase)      \\
                       &     &
                          &    {\boldmath $\sim -15\%$ \ -- \ $+15\%$}  \\
                    &  &  &  (w/ soft isospin phases) \\
\hline
{\boldmath $b\to ss\overline s$}
                          &    $B\to K\phi,\ K^{*}\phi$
                          &    ${\cal O}(10^{-5})$
                          &    ${\cal O}(1\%)$          \\
\{pure penguin\}
                          &           &         &        \\
\hline
{\boldmath $b\to ds\overline s$}
                          &    $B^+\to K^+\bar K^{0*},\ K^+K_S$
                          &    ${\cal O}(10^{-6})$
                          &    {\boldmath ${\cal O}(${\bf few} -- $30)\%$}\\
\{pure $b\to d$ penguin\}
                          &           &         &        \\
\hline
{\boldmath $b\to dc\overline c$}
                          &    $B^+\to D^{+\left(*\right)}
\overline{D}^{0\left(*\right)}$
                          &    ${\cal O}(0.02\% - 0.11\%)$
                          &    ${\cal O}(1\%)$   \\
\{tree $\oplus$ penguin\}
                          &
                          &
                          &    (perturbative phase)      \\
\hline
\multicolumn{4}{@{}p{120mm}}{Adapted from: J.-M. G\' erard and W.S. Hou,
                                          \cite{GH91a,GH91b}, and
                                                          I. Dunietz and W.S.
Hou, \cite{DH94}.}
\end{tabular*}
\end{table*}

For the evaluation of $a_{CP}$,
which in addition depends on absorptive parts,
there are two further basic questions/problems.
\begin{enumerate}
\item What is $q^2$ of virtual gluon?

For inclusive processes, $q^2$ is one of the phase space integration
variables, and it is a fair approximation to assume perturbative QCD.
For exclusive processes, the meson
$B(b\bar q^\prime)$ decays into $s\bar q q\bar q^\prime$,
and typically $s\bar q$ forms a $K$, $K^*$, $\phi$ {\it etc.},
while  $q\bar q^\prime$ forms $\pi$, $\rho$, $K$, $K^*$, {\it etc.}.
Then $q^2$ would nominally correspond to the invariant mass squared
of the $q\bar q$ pair. It is not clear what it actually means.
On one hand, one may argue that the short distance rescattering
should be summed over.  One therefore may be tempted to take
$a_{CP}$ from the inclusive result of Table 1.
On the other hand, one observes that
the $q\bar q$ pair has to be more or less back-to-back
and of high momentum to end up in a two-body meson final state.
Furthermore, the spectator $\bar q^\prime$ has to receive
a kick to be sped up as well.
{}From these lines of arguments, G\' erard and Hou argue that
the effective $q^2$ should be in the range
$m_b^2/4 < q^2 < m_b^2/2$ \cite{GH89}.

The issue remains unsettled, and is symptomatic of the main problem
for direct $CP$ violation observables as opposed to mixing related ones:
one does not have good control of the participating absorptive parts.

There is one recent development that is worthy of note, however.
Akhoury, Sterman and Yao \cite{ASY93} have recently applied perturbative QCD
with Sudakov suppression for a serious, ``first principle'' evaluation
of $B\to \pi\pi$ decay rate. Working further along this line
may shed some light on the issue of absorptive parts and
effective $q^2$ value for two body decays.

\item Soft phases for $b\to su\bar u$ exclusive modes.

The $b\to su\bar u$ type of modes are special in that they have both
tree and penguin contributions.
The former has isospin 0 and 1 components while
the latter is purely isospin 0 (in the transition).
Because of presence of two isospin amplitudes,
unlike the pure penguin case, {\it soft} strong interaction phases
are very important at the exclusive level \cite{GH91a}.
At  5 GeV center of mass energy,
these soft strong rescattering phases are not known
and it is not clear whether they would ever be measurable.
\end{enumerate}

With these comments, we present some numerical results
in Table 2 as illustration for the range of ``predictions",
and thereby the uncertainties involved.
The $b\to dc\bar c$ entry would be discussed in next section.

When comparing with the standard mixing-dependent $B\to \psi K_S$ mode,
some easily overlooked prejudices should be removed.
The branching ratio of  $\mbox{few}\times 10^{-4}$
and advertised asymmetries of order $10-50\%$ both seem
large in comparison.
However, when the branching of $\psi\to \ell^+\ell^-$ and
$K_S\to \pi^+\pi^-$ are folded in,
the usable branching ratio is again
of order $10^{-5}$. At the same time,
to observe $a_{CP} \sim 10-50\%$ one
needs to reconstruct {\it both} vertices (double vertices)
and tagging the $b$ flavor.

Let us conclude the discussion of exclusive
charmless $B$ decays with the following observations:
\begin{itemize}
\item $b\to ds\bar s$ pure penguins
({\it e.g.} $B^+\to K^+K_S,\ K^+\bar K^{0*}$)

Despite having the smallest branching ratios,
we believe these are the most promising direct $CP$ violation
modes in the Standard Model \cite{GH91b}.
With $\mbox{BR} \sim \mbox{few}\times 10^{-6}$
and $a_{CP}$ potentially as large as $20 - 30 \%$,
{\boldmath
${\cal N}_B \ge \mbox{\bf few}\times 10^7$ could suffice for each mode}
(here ${\cal N}_B$ stands for number of $B$'s needed).
Moreover, since pure penguin modes have only one
isospin component, one can sum over modes without
suffering cancellations. If the inclusive results of Table 1
can be of any guide, {\it theoretically, with $\eta \sim 0.3$,
a few $\times 10^6$ $B$'s could already suffice!}
One should therefore put some premium on searching for these
very rare loop-induced decays.

\item $b\to su\bar u$ ({\it e.g.} $B^+\to K^+\pi^0$, $B^0\to K^+\pi^-$)

The branching ratio is at the $10^{-5}$ level.
CLEO had the first glimpse of these modes in 1993,
although they are not yet separable from $\pi\pi$.
The key point here is that difference in
soft phases between the two isospin amplitudes may make
$\vert a_{CP}\vert$ as large as $15\%$ \cite{GH91a}.
Thus, again, ${\cal N}_B \ge $ a few $\times 10^7$
could suffice to uncover $CP$ violating asymmetries
for a particular mode.
However, the same reason that soft phases enter,
if they are sizable at all, guarantees large cancellations
when one atempts at accumulating different modes.
That is, if the asymmetries per mode can be as large as
$10\%$, it is necessarily due to soft phase differences,
but then $a_{CP}$ would necessarily
fluctuate from mode to mode.
It may also happen that the two isospin components
receive similar soft phases, then one
would again have per cent level asymmetries and
${\cal N}_B \ge 10^9$.
One should therefore try hard for every possible mode,
but attempts at summing over different modes would be in vain.

\item $b\to ss\bar s$ pure penguin ({\it e.g.} $B^+\to K^+\phi$)

With $\mbox{BR} \sim 10^{-5}$ and $a_{CP} \le 1\%$,
we expect ${\cal N}_B \ge 10^9$ and Standard Model
predicts no realistically observable $CP$ asymmetry
in these modes.
However, this is a consequence of small 3 generation $CP$
violating phase that enters via the penguins.
In case 3 generation unitarity is violated in
any significant way, new physics effects could show up
as $a_{CP} > 1\%$.

\item Experimental relevance: Discovery Modes

Note that all of the above modes lead to final states
constituting of 2 -- 4 $K$ or $\pi$ mesons (or more).
That is,
\begin{equation}
\left\{     \begin{array}{l}
              \pi\pi\pi,\ \pi\pi\pi\pi,   \\
              K\pi,\ K\pi\pi,\ K\pi\pi\pi,   \\
              K\overline{K},\ K\overline{K}\pi,\ K\overline{K}\pi\pi,   \\
              KK\overline{K},\ KK\overline{K}\pi,\ \mbox{\it etc.},
            \end{array}
\right.
\end{equation}
where $\pi\pi$ may form $K_S$ or $\rho$, $K\pi$ may form $K^*$,
while $K\overline{K}$ may be $\phi$, {\it etc.}
Direct, non-resonant decays into final states given in eq. 7
may be prominent. However, effective two body modes
may be more realistic in suppressing combinatoric background.

Hence, to study direct $CP$ violation effects, one needs
good particle identification, {\it i.e.}
$K/\pi$ separation for the full momentum range.
For asymmetric $B$ Factories proposed at KEK and SLAC,
this is up to momentum of 4 GeV/c or so.
One should remember that, although good $K/\pi$ separation
is not needed for the most famous $B\to \psi K_S$ mode,
they are again crucially needed for other  types
of mixing-dependent modes such as $B_d^0\to \pi^+\pi^-$ (suppression
of $K\pi$ penguin). They are likewise needed for the method
proposed by Gronau and Wyler \cite{GW91}
to measure $\gamma\ (\phi_2)$ by studying $B\to DK$.
\end{itemize}

\section{Another Kind of Charmless: $b\to dc\bar c$ [7]}

Traditionally, charmless final states refer to
those with no charm particles whatsoever.
A second type of ``charmless" final state would
be those that {\it have} charm particles,
but coming in conjugate pairs so total charm quantum number
$C = 0$. The strong interaction clearly can reshuffle
these two classes of charmless final states.
Indeed, we have seen previously that
$b\longrightarrow sc\bar c\to sg^* \to su\bar u$
rescattering to be a source of absorptive part.
Here, we have the opposite.
The tree level (dispersive) $b\to dc\bar c$ amplitude
can interfer with
\begin{equation}
\overbrace{b\longrightarrow du\bar u}^{\mbox{ Weak}}
\underbrace{\to dg^*
                     \to}_{\mbox{Strong}}
dc\bar c
\end{equation}
which is absorptive. The physical  $du\bar u$ final state
could rescatter inelastically via strong interactions.
We pick $b\to d$ penguins since, as we have seen earlier,
they have more favorable KM phase combinations.

\subsection{Illustrative estimate}

Let us make a schematic estimate of the ensuing
branching ratios and rate asymmetries.
We have $B^\pm$ decay in mind, since
for $B^0$ decay, the $d\bar c c\bar d$ final state is self-conjugate.
Note that there is only one single isospin amplitude.
The tree level $b\to dc\bar c$ amplitude dominates the decay,
and has the KM factor
\begin{equation}
v_c = V_{cd}^* V_{cb} \simeq -A\lambda^3,
\end{equation}
where we have used the Wolfenstein parametrization.
Taking the remaining tree amplitude as unity (and real),
we need only the rescattering, therefore imaginary, amplitude
of process given in eq. 8. This has the KM factor
\begin{equation}
v_u = V_{ud}^* V_{ub} \simeq A(\rho-i\eta)\lambda^3.
\end{equation}
The remainder of the amplitude is ${\cal O}(\alpha_s)$
$u\bar u \to g^* \to c\bar c$ rescattering. Since
$q^2$ is way above $c\bar c$ threshold,
the amplitude can be taken as purely absorptive
and therefore purely imaginary.
Altogether one has
$-i\, \alpha_s /4\pi$.
Ignoring operator differences,
the amplitude for $B^-$ decay is
\begin{equation}
{\cal M} \propto -A\lambda^3\left[1 + (\rho-i\eta)\,
                                      {\alpha_s\over 4\pi}\, i \right],
\end{equation}
while for $B^+$ decay, $\eta$-term changes sign.
We thus find
\begin{eqnarray}
 \mbox{BR}(b\to dc\bar c) \simeq
       \left\vert{V_{cd} \over V_{cs}}\right\vert^2
                                    \mbox{BR}(b\to sc\bar c),      \\
                                                       \nonumber
\\
a_{CP} \simeq \mbox{$\vert {\cal M}\vert^2 - \vert \overline{\cal M}\vert^2
                \over
                \vert {\cal M}\vert^2 + \vert \overline{\cal M}\vert^2$}
        \simeq {\alpha_s \over 4\pi}\, (2\eta) \sim 1\%,
\end{eqnarray}
where we have taken $\alpha_s/4\pi \simeq 0.02$ and
$2\eta \sim 0.5$.
The branching ratio relation of eq. 12 should
hold for both inclusive and exclusive modes.

We emphasize that the rate is tree dominant.
Since $b\to sc\bar c$ modes have already been seen,
the BR estimates are rather reliable.
The $a_{CP}$ estimate is also reliable since
one is certainly in the perturbative domain.
Note that although BR is much larger for $b\to sc\bar c$,
the accompanying $a_{CP}$ is smaller by $\lambda^2$,
of order $0.05\%$, way too small for sake of observation.
Exclusively, one needs to study, for example,
$B^-\to D^{0\left(*\right)} D^{-\left(*\right)}$ {\it vs.}
$B^+\to \overline{D}^{0\left(*\right)} D^{+\left(*\right)}$.
Scaling from observed
$B\to D^{\left(*\right)} \overline{D}_s^{\left(*\right)}$
branching ratios, one finds
$\mbox{BR}(B\to D^{\left(*\right)} \overline{D}^{\left(*\right)})$
ranges from $0.02\%$ to $0.11\%$, which agrees with model estimates.
The BR and asymmetry estimates are listed also in Table 2.
The crucial point, however, is probably distinguishing
$D$ from $D_s$, since full reconstruction of both $D$ mesons
would be rather punishing.

\subsection{Inclusive: further illustration}

We could use the inclusive results of Sec. 2.1
to further illustrate the point.
The theoretical estimates for inclusive $b\to dq\bar q$ ($q\neq c$)
is more reliable than the correspondings exclusive modes.
Since the strong interaction mediates
$q\bar q \Longleftrightarrow c\bar c$ rescattering,
by $CPT$ and unitarity arguments
\begin{equation}
   \Delta\Gamma_{b\to dq\bar q} =
 - \Delta\Gamma_{b\to dc\bar c}
\end{equation}
that is, the increase(decrease) in inclusive $b\to dq\bar q$ rate
corresponds to decrease(increase) in $b\to dc\bar c$ rate.
Strong interaction just reshuffles among the modes.
Thus, from Table 1,
\begin{eqnarray}
   \Delta \mbox{BR}_{q\bar q}  &  =  &  - \Delta \mbox{BR}_{c\bar c}  \nonumber
 \\
                                                      &  =  &  \mbox{BR}(b\to
dq\bar q)\cdot a(b\to dq\bar q)

                                          \nonumber   \\
                                                      &  \simeq   &  0.8\%
\cdot (-0.8\%),
\end{eqnarray}
where, in the last step we have assumed $\eta = 0.3$
rather than the value of 0.15 used in Table 1.
{}From
\begin{equation}
\mbox{BR}(b\to dc\bar c) \simeq \lambda^2 \mbox{BR}(b\to sc\bar c) \sim 0.75\%,
\end{equation}
we find
\begin{equation}
a(b\to dc\bar c) \simeq {0.64\times 10^{-4}\over 0.75\times 10^{-2}}
                 \simeq 0.85\%.
\end{equation}
These results are also listed in Table 1.
Note that  the asymmetry is basically unaffected by two-loop effects.
Thus, at the $3\sigma$ level, in principle
\begin{equation}
{\cal N}_B \sim 9\, \mbox{BR}^{-1}\, a_{CP}^{-2} \ge 2\times 10^7
\end{equation}
$B$ mesons would suffice for uncovering $CP$ violation
in $b\to dc\bar c$ modes. The question is whether one
could develop  methods for inclusive study.  Again, one has to
eliminate $b\to sc\bar c$ ``background", which has vanishingly small
asymmetry but  BR at $15\%$ level.

Note that a second source for $CP$ violation
is possible: $b\bar u$ annihilates into $d\bar u$,
pulling out $c\bar c$ from QCD vacuum.
The annihilation diagram, however, cannot be reliably calculated,
and is likely very small.

The cleanliness of $b\to (d\mbox{\ or\ }s)c\bar c$
for $CP$ violation studies was already mentioned by BSS,
was followed up by Chau and Cheng \cite{CC87},
and can be found in a recent work by Simma and Wyler \cite{SW91}.
However, these modes have not been stressed in the literature
since the late 1980's.
Somehow, the uncovering of subtleties in
$b\to (d\mbox{\ or\ }s)q\bar q$ modes has had the side effect of
suppressing the  $b\to (d\mbox{\ or\ }s)c\bar c$  modes
from public consciousness.
It needs to be re-emphasized so experimenters can be prepared
when $B$ Factories turn on.
If one reconstructs the $D$'s, with good tracking and
momentum resolution, one has less demands on particle identification.
Perhaps one can then profit from the high production cross section
at hadronic machines.

\end{document}